# Failure of Delayed Feedback Deep Brain Stimulation for Intermittent Pathological Synchronization in Parkinson's Disease

Andrey Dovzhenok[1], Choongseok Park[1], Robert M. Worth[1,2] and Leonid L. Rubchinsky[1,3]

[1] Department of Mathematical Sciences and Center for Mathematical Biosciences, Indiana University Purdue University Indianapolis, Indianapolis, Indiana, United States of America

[2] Department of Neurosurgery, Indiana University School of Medicine, Indianapolis, Indiana, United States of America

[3] Stark Neurosciences Research Institute, Indiana University School of Medicine, Indianapolis, Indiana, United States of America

**Corresponding author:**

Andrey Dovzhenok

Department of Mathematical Sciences

University of Cincinnati

2815 Commons Way, Cincinnati, OH 45221 USA

E-mail: andrey.dovzhenok@uc.edu;   Phone: 513-556-4160

# Abstract

Suppression of excessively synchronous beta-band oscillatory activity in the brain is believed to suppress hypokinetic motor symptoms of Parkinson's disease. Recently, a lot of interest has been devoted to desynchronizing delayed feedback deep brain stimulation (DBS). This type of synchrony control was shown to destabilize the synchronized state in networks of simple model oscillators as well as in networks of coupled model neurons. However, the dynamics of the neural activity in Parkinson's disease exhibits complex intermittent synchronous patterns, far from the idealized synchronous dynamics used to study the delayed feedback stimulation. This study explores the action of delayed feedback stimulation on partially synchronized oscillatory dynamics, similar to what one observes experimentally in parkinsonian patients. We employ a computational model of the basal ganglia networks which reproduces experimentally observed fine temporal structure of the synchronous dynamics. When the parameters of our model are such that the synchrony is unphysiologically strong, the feedback exerts a desynchronizing action. However, when the network is tuned to reproduce the highly variable temporal patterns observed experimentally, the same kind of delayed feedback may actually increase the synchrony. As network parameters are changed from the range which produces complete synchrony to those favoring less synchronous dynamics, desynchronizing delayed feedback may gradually turn into synchronizing stimulation. This suggests that delayed feedback DBS in Parkinson's disease may boost rather than suppress synchronization and is unlikely to be clinically successful. The study also indicates that delayed feedback stimulation may not necessarily exhibit a desynchronization effect when acting on a physiologically realistic partially synchronous dynamics, and provides an example of how to estimate the stimulation effect.

# Introduction

Deep brain stimulation (DBS) entails the delivery of a stimulation signal to subcortical structures via implanted electrodes. DBS has received a lot of attention as a therapeutic procedure in various neurological and neuropsychiatric disorders [1]. DBS of different targets in the basal ganglia-thalamocortical loop is used to treat symptoms of Parkinson's disease (PD) and other motor disorders [2]; e.g. the subthalamic nucleus (STN) is a standard anatomical target for DBS in PD.

The hypokinetic symptoms of PD have been related to excessive beta-band oscillations and synchrony in the basal ganglia and other structures [3-5]. Thus DBS effectiveness has been linked to the destruction of this pathological rhythmicity by reducing the bursting, oscillations and synchronization in the beta-band and increasing regularity and synchrony in the high-frequency band [6-12].

However, standard DBS, while clinically effective, does not completely restore motor function and has substantial side effects, which may be related to its strong stimuli and "one size fits all" approach. Standard DBS is associated with a variety of adverse effects such as dyskinesia, paraesthesia, dysarthria and gait disturbances [13]. Non-motor adverse effects (mania, impulsivity, depression, various cognitive alterations, suicidal behavior, etc.) are also a problem [14]. They can arise due to current spread to adjacent structures and due to the fact that associative, limbic and motor circuits, although traditionally viewed as largely parallel in the basal ganglia, are not completely independent [15].

These considerations lead to a strong interest in new DBS algorithms. Ideally, stimulation waveforms should have small amplitudes and should be targeted specifically to destruction of the pathological activity which results in the primary symptoms. Low amplitudes of stimulation will also save battery life, reducing the need for battery-replacement surgeries.

One method which has received a lot of attention recently and has appeared to be very promising is delayed feedback. This elegant feedback control scheme rendered the synchronized state in an ensemble of all-to-all coupled oscillators unstable [16,17]. In the limit of a large number of oscillators, the amplitude of feedback signals vanishes, which makes it especially attractive. This control scheme was modified into a more realistic setting by using delayed feedback based on a mean field signal (a proxy for easy-to-record local field potentials, LFP) in order to cancel the effect of coupling and desynchronize ensembles of coupled oscillators [18,19]. Subsequent studies provided further computational evidence for the ability of delayed feedback to destabilize a synchronized state (e.g., [20-24]).

Therefore, the delayed feedback desynchronization algorithm appears to be quite robust. However, in spite of these advances and in spite of hardware availability [25], to our knowledge, this strategy has never been implemented in patients. There may be many reasons why the considered desynchronization technique has resisted effective clinical realization. The goal of this paper is to explore the action of delayed feedback DBS in a realistically partially-synchronous network. We conjecture that a complex origin of partially synchronous neural dynamics in parkinsonian brain may be a substantial obstacle to the implementation of delayed feedback desynchronization.

To study this problem, we employ a computational model of the basal ganglia network which successfully reproduces experimentally recorded neural activity [26]. The synchronous activity in the PD brain is very intermittent [27-29]. The model in [26] is based on the membrane properties of the basal ganglia cells and is tuned in such a way as to reproduce not only the average synchrony levels, but also the temporal patterns of the synchronous dynamics seen in human experimental data. In the language of dynamical systems theory, this model realistically describes the dynamics not only in the vicinity of the synchronized state, but also in other parts of the phase space, ensuring more similarity between the model and the experimental system [30]. In contrast, earlier studies used neural oscillators in a fully synchronized regime.

We will investigate the action of delayed stimulation as we vary network parameters to go from completely synchronized dynamics to more realistic intermittent synchrony. As a result we can see how delayed feedback DBS is performing in a setting whose dynamics is more faithful to that seen in PD patients.

## Methods

*Model Network*

We used the basal ganglia model studied in [26]. This model is based on [31], but its dopamine-modulated parameters are tuned to reproduce experimentally recorded data. The model consists of two arrays: an array of 10 Globus Pallidus externus (GPe) model neurons and an array of 10 STN model neurons; each array assumes a circular structure with respect to inhibitory connections from GPe to STN neurons (for example, $10^{th}$ GPe neuron inhibits $9^{th}$, $10^{th}$ and 1st STN neurons). Each GPe neuron receives synaptic input from one STN neuron, while each STN neuron receives inputs from the same neuron to which it transmits as well as from two of its neighbors (Figure 1). While the model is clearly limited in many ways and does not incorporate other brain structures beyond STN and GPe, it is based on experimental anatomical and physiological data and captures the rich experimentally recorded repertoire of PD rhythmicity [26]. In addition, it appears to adequately reproduce the experimentally studied mechanisms of this rhythmicity resulting from sequences of recurrent excitation and inhibition in STN-GPe networks [32,33].

Both subthalamic and pallidal neuron models are described by conductance-based (Hodgkin-Huxley like) formalism, with channel properties recovered from experiment [31]. The model includes leak current, fast spike-producing K and Na currents, low threshold T-type and high-threshold $Ca^{2+}$-currents, and $Ca^{2+}$-activated voltage-independent afterhyperpolarization (AHP) $K^+$-current in the current balance equation:

$$C\frac{dV}{dt} = -I_L - I_K - I_{Na} - I_T - I_{Ca} - I_{AHP} - I_{syn} + I_{app}$$

where $I_L = g_L \cdot (V - V_L)$, $I_K = g_K n^4 \cdot (V - V_K)$, $I_{Na} = g_{Na} m_\infty^3(V) h \cdot (V - V_{Na})$, $I_T = g_T a_\infty^3(V) b_\infty^2(r)(V - V_{Ca})$, $I_{Ca} = g_{Ca} s_\infty^2(V)(V - V_{Ca})$, $I_{AHP} = g_{AHP}([Ca]/([Ca]+k_1))(V - V_K)$. The intracellular concentration of calcium is described by $d[Ca]/dt = \varepsilon(-I_{Ca} - I_T - k_{Ca}[Ca])$. The gating variables *n, h* and *r* obey 1st-order kinetic: $dx/dt = (x_\infty(V) - x)/\tau_x(V)$. Fast activation variables *m, a* and *s* are assumed to be instantaneous with voltage-dependent activation functions $m_\infty(V)$, $a_\infty(V)$ and $s_\infty(V)$, correspondingly. Synaptic current is $I_{syn} = g_{syn}(V - V_{syn})\sum_j s_j$, where the sum is taken over the presynaptic neurons from which there are incoming connections to the given cell. The synaptic variable $s_j$ obeys

$$ds_j/dt = \alpha H_\infty(V_{presyn} - \theta_g)(1 - s_j) - \beta s_j.$$

While both STN and GPe neurons are described by the same kind of equations, the parameters of these equations are different, reflecting the difference in the biophysical properties of their neuronal membranes. GPe and STN parameters follow [26].

*Stimulation Setup*

The stimulation setup is given in Figure 2. Time-delayed feedback is used for stimulation current following the ideas of [16]–[19]. This type of feedback reliably disrupts correlated activity in a model of synaptically-coupled neuronal systems.

The feedback signal is constructed by computing LFP, following the rationale in [26]. Hence, STN LFP at the $j^{th}$ neuron is:

$$X_j(t) = -i_{syn,j} - w_1(i_{syn,j-1} + i_{syn,j+1}) - w_2(i_{syn,j-2} + i_{syn,j+2}) + I_{stim,j}$$

where $i_{syn,j}$ is the total synaptic current coming to the neuron $j$; $w_1$ and $w_2$ are weights representing the attenuation of the field with the distance. We set the weights $w_1$ and $w_2$ to several values from [0, 0.4]. We consider $w_1 > w_2$ to account for the attenuation of the signal with $w_2 = 0.1$ or zero. However, for both choices of $w_2$ results were qualitatively similar.

The model LFP is measured at the same site at which stimulation is applied, and so, the stimulation current $I_{stim,j}$ is added to $X_j(t)$. The resulting signal is then filtered using the damped harmonic oscillator as suggested in [20]:

$$\ddot{x}_j + \alpha_f \dot{x}_j + \omega^2 x_j = X_j(t),$$

where $\omega = 2\pi/T$ and $T$ is the mean period of bursting in the model network without stimulation. Parameter $\alpha_f = \omega$ determines the band pass properties of the filter. To compensate for a phase shift introduced by filtering, the output of the harmonic oscillator is delayed by the value $\tau_s$ of the shift [19]: $\tilde{x}_j(t) = x_j(t - \tau_s)$. The period $T$ of bursting activity in the parameter region investigated in the manuscript does not vary much. Therefore, time-delayed differential feedback is computed as $\tilde{x}_j(t - \tau) - \tilde{x}_j(t)$, where $\tau = T/2$. Finally, the stimulation signal might become very strong and present danger to neuronal cells; so it is desirable to bound the stimulation signal. Here, we use a nonlinear transformation of the filtered signal that keeps the stimulation current strength bounded. The feedback stimulation current at the $j^{th}$ STN neuron is then obtained as:

$$I_{stim,j} = 2K_f \left(1/\left(1 + \exp\left(-C_1\left(\tilde{x}_j(t-\tau) - \tilde{x}_j(t)\right)/14\right)\right) - 0.5\right) \\ + w_1(I_{stim,j-1} + I_{stim,j+1}) + w_2(I_{stim,j-2} + I_{stim,j+2})$$

where $C_1$ is a normalization parameter of the stimulation signal before the nonlinear transformation, the factor /14 corresponds to the nonlinear transformation and defines the slope of the bounded signal at the point where the delayed signal vanishes (and hence the overall shape of the stimulation signal) and was chosen such that the general shape of the unbounded signal was not excessively altered by the bounding function and $K_f$ is an amplifying step of the stimulation setup after application of nonlinear transformation (see Figure 2).

The STN-GPe model network [26] that we utilize in the current study presents an example of an inhibitory-excitatory network. Anti-phase oscillations are common in this type of network. Therefore, below we consider different spatial electrode setups to rule out the possibility that only some particular stimulation arrangements are effective in suppression of synchrony (for example, only adjacent or only non-adjacent electrode setups are effective).

In our simulations, we administer stimulation current through one to three electrodes and 30% to 90% of the model STN network was affected by stimulation depending on a particular electrode set-up. Electrodes are placed in the following arrangements: a single electrode, two electrodes positioned near adjacent STN neurons, three electrode positioned near adjacent neurons, two electrodes positioned near nonadjacent neurons (stimulation electrodes are placed near STN neurons $j$ and $j+2$), and three electrodes positioned near nonadjacent neurons (electrodes are placed near STN neurons $j-2$, $j$, and $j+2$). This gives a total of five different electrode placement arrangements that we investigate in this paper.

In numerical simulations, stimulation feedback was switched on 1 s after the start of simulations. A second later, the data was saved for 5 s and was subjected to the analysis (see below). The model network equations were numerically solved with XPP software (Bard Ermentrout, University of Pittsburg, http://www.math.pitt.edu/~bard/xpp/xpp.html).

*Network's dynamics and estimation of its synchrony*

The model network without stimulation was analyzed earlier in [26] in the two-dimensional parameter space of ($g_{syn}$, $I_{app}$) – the strength of GPe to STN synaptic connections and the applied current to the GPe neurons ($I_{app}$ represents synaptic input from striatum to pallidum). The choice of these parameters was grounded in the following considerations [26,29]. Both of these parameters (essentially, synaptic strengths) are affected by dopamine. In PD, nigral dopaminergic cells degenerate, thus depriving these synaptic connections of dopaminergic modulation. Larger values of $g_{syn}$ and smaller values of $I_{app}$ would correspond to a PD-like state. In numerical experiments this would lead to more synchronous dynamics.

To estimate the amount of synchrony in the network principal component analysis (PCA) was used following [26]. PCA components give a measure of overall, global coherence in the network and as such are very convenient in the present study. Moreover, we compare our simulation results to results of [26] that successfully utilized PCA to measure the level of synchrony in the model network. The slow variable *r* from each of the STN neurons was used for the analysis (we choose the slow variable because beta-band synchrony here is essentially a synchrony of bursting). We look at the number of principal components capturing 80% of the variation.

The dynamics of the network without stimulation is presented in Figure 3. The right lower corner of the network is a synchronized state, while the left upper corner is nonsynchronized state. The dashed contours in the figure indicate the parameter domain where the dynamics of the model network exhibits synchronous patterns similar to what is experimentally observed not only in average synchrony level, but also in the fine temporal structure of synchrony [26]. This area of parkinsonian dynamics is on the boundary between the nonsynchronous state and an unrealistic strongly synchronous state. Given the location of the realistic firing patterns we confined our simulations to a smaller domain, which still captures the main types of dynamics (dotted contour, Figure 3). The effect of the stimulation on the degree of synchrony was measured by the change in the number of principal components in the network with stimulation vs. without stimulation. Time-series analysis was done in MATLAB (Mathworks, Natick, MA).

## Results

Depending on the values of $g_{syn}$ and $I_{app}$ the STN-GPe model network may exhibit three characteristic types of collective behavior: irregular activity, strongly correlated spiking, or an intermittent synchrony regime on the boundary between the former two [26]. The intermittent activity in the model possesses the same temporal synchronization pattern as recorded from STN neurons in patients with PD [28]. Therefore, when measuring the effect of proposed feedback stimulation we were particularly interested in how delayed feedback stimulation acted on this realistically intermittent weak synchrony.

*Examples of synchronizing and desynchronizing action of delayed feedback stimulation*

An example of the action of feedback on strongly synchronous dynamics is given in Figure 4A. It can be seen (Figure 4A) that the stimulation leads to a reduction in synchrony and more uncorrelated dynamics, i.e. the phase locking between stimulated neurons is broken by the delayed feedback and this result extends to the whole STN network. Interestingly, there appears to be a delay of about 500 ms (Figure 4A) before the stimulation signal produces a desynchronizing action in the model network, and this effect diminishes with stronger stimulation strength (not shown). This delay in desynchronization is most likely due to the type of synchronization regime in which network STN cells form two clusters of neurons oscillating in anti-phase. In the beginning, delayed feedback current applied in the two stimulated cells drives them in-phase with each other but also in-phase to the cells next to them in the network (that oscillate in anti-phase in stimulation-free ensemble). Hence, desynchronizing action is likely achieved by breaking synchrony between the two clusters in the network.

On the contrary, in the intermittent regime the same delayed feedback stimulation results in no apparent change in synchronization for moderate stimulation strengths, while stronger stimulation, in fact, leads to increased synchronization among STN neurons (Figure 4B). While Figure 4B illustrates the dynamics of two neurons in the network, the synchronizing effect of the "desynchronizing" feedback stimulation is confirmed by the decrease in the number of principal components for the whole network as can be seen in Figure 5.

*Delayed feedback effects on networks with different synchrony levels*

To study these phenomena systematically, we consider the dynamics of the network in the two-dimensional space of parameters $g_{syn}$ and $I_{app}$ and vary strength of the feedback stimulation. We start by setting $I_{app}$ at some intermediate value that, depending on the parameter $g_{syn}$, produces either intermittent synchrony or strongly correlated activity. Figure 5 depicts the change in the number of principal components in the network stimulated with delayed feedback compared to the network without stimulation. Here, the increase in stimulation strength leads to decrease in synchrony in the network (indicated by the increase in the number of principal components) when the synaptic parameter $g_{syn}$ corresponds to the strongly correlated activity without stimulation (see Figure 3). However, the model network which is in an intermittent synchronization regime before stimulation (see Figure 3) shows no positive change in the number of principal components and eventually becomes more synchronous with stronger stimulation current. This is highlighted by the decrease in the number of principal components with higher values of $K_f$. Thus there is a marked difference in a trend: as $g_{syn}$ decreases to produce less coherent pre-stimulation dynamics, the increase in the stimulation strength leads to more rather than less synchronized dynamics.

The results for other types of spatial arrangement of stimulation electrodes are presented in Figure 6. One can see that some stimulation set-ups may lead to desynchronizing effect even for moderate values of $g_{syn}$, however, there are nearby values of $g_{syn}$ which yield no improvement in desynchronization.

Similar phenomena were observed by us for other values of $I_{app}$. Therefore, for a systematic study of these phenomena we will vary both control parameters ($g_{syn}$ and $I_{app}$) in the model network to span a large repertoire of synchronized behavior and to include synchrony patterns similar to experimentally observed ones. To find the largest possible desynchronizing effect of the delayed feedback, we consider the maximum increase in the number of principal components, that is, the maximum desynchronization effect, in the two-parameter plane $g_{syn}$-$I_{app}$ obtained over the full range of tested stimulation strengths for the electrode arrangements from Figure 5 (Figure 7). The only consistent improvement in desynchronization was made in the region of strongly correlated activity (see Figure 3). For the parameter values corresponding to uncorrelated activity and intermittent synchrony desynchronization of the network was not usually achieved. On the contrary, as Figure 5 shows, stronger delayed feedback stimulation at these parameter values frequently leads to stronger correlation and overall more synchronous dynamics.

Similar to Figure 7, the effect of spatial electrode arrangements considered in Figure 6 is summarized in Figure 8. Therefore, while the delayed feedback stimulation produces reliable synchrony suppression in the case of strongly correlated activity, it frequently fails to destroy synchronized activity in networks with intermittent synchrony regimes like those observed in PD patients.

## Discussion

*Potential limitations of the modeling*

The modeling approach used here does not include many factors involved with physiology of PD. The real mechanisms of the generation of synchronized beta-band oscillations may be much more complicated. However, the model reproduces the experimentally observed synchrony patterns [26], and thus it appears to be dynamically adequate for studying the real basal ganglia circuits in PD. There is an equivalence of the phase spaces of the model and of the real dynamics not only in the vicinity of the synchronization manifold, but in other areas of the phase space. This is important because the overall synchrony is not strong and substantial fraction of time is spent in those areas of the phase space.

The model LFP here involves 30% or 50% of STN network (in line with [26]) and only 30% or 50% of STN are stimulated by a single electrode in the model. However electrode arrangements affecting almost all STN neurons were also studied and no qualitative difference from other arrangements was found.

The number of neurons in the model is relatively small and small-size effects do exist. However the twenty-neuron model used here was previously validated with experimental data in [26]. Moreover, dynamics in the networks of many elements is based on the same mechanisms described here [34,35] but computational times will grow enormously. Similar to the stimulation of STN in a real patient, stimulation current affects a substantial part of the model STN. The results obtained are in agreement with what we have previously found for large networks where stimulation desynchronizes the neuronal ensemble in the strong coupling limit. Small size effects manifest themselves in the model in that the synchronization suppression is good, but not perfect. Therefore, we expect our results to hold in a larger network as well.

The small size of the model network is also a reason why stimulation current does not vanish in a desynchronized state. Desynchronization with a completely vanishing stimulation (i.e. when the control signal is almost zero in the desynchronized state) is possible only in the limit of a very large number of oscillators. However, this should not negate the observations of this study. We follow the change in the degree of synchrony in the network, and what is important is the direction of the effect (more or less synchronizing action of stimulation).

Our results here are understandable because all the simulations done to date as well as the underlying delayed feedback theory developed were for the case when synchronization is relatively strong. In parkinsonian brain, however, synchronization is highly intermittent [27-29] with distinctive temporal patterns so that the theory developed may not apply.

Modeling studies suggest that the structure and parameters of the feedback stimulation affect the efficiency of desynchronization. For example, computing the mean field from a group of elements not completely coincident with the group of stimulated elements made the domain of existence of desynchronization smaller [36]. This may be of potential relevance to the subthalamic nucleus, because the mean field is likely to be generated by pallidal synaptic activity and is represented in such a way in the model utilized here. Also, for moderate strengths of the feedback loop, some nonlinear arrangements of delayed feedback stimulation may exert a synchronizing effect; however the desynchronizing effect occurs for stronger feedback stimulation [21]. But these issues are unlikely to vitiate the major result of our study. We had no problem in obtaining desynchronization in a network which is fully synchronous to begin with. However as parameters of the network are gradually changed in such a way as to obtain experimentally realistic, partially synchronous firing patterns the "desynchronizing" feedback gradually loses the ability to decrease synchrony strength in the system and, in fact, eventually increases the synchrony level. We varied the strength of the feedback for each of the parameter sets of the model network to find the optimal stimulation characteristics, but it did not affected the general outcome. We do not completely exclude the possibility that *some* feedback control may potentially decrease synchrony of a partially synchronized dynamics in PD basal ganglia. However our results indicate that *the same* delayed feedback stimulation that desynchronizes complete synchrony may actually increase synchrony strength in a partially synchronized regime.

*Conclusions*

Our results indicate that delayed feedback is more likely to increase synchrony in the basal ganglia of PD patients rather than to suppress it. This warrants investigation of other DBS techniques. For example, coordinated resetting (e.g., [37]) may be an effective desynchronizer (it may be also beneficial due to the improvement in thalamocortical relay function, [24]). However, unlike delayed feedback stimulation it does not vanish in the limit of a large number of oscillators.

Another non-vanishing, but potentially efficient technique is based on the optimization of the stimulation waveforms [38], where the stimulation signals are drawn from a broad class of waveforms and optimized by genetic search algorithms. An emerging Kalman filtering approach also appears promising [39].

The other important implication of the present study extends beyond the context of DBS in PD. Our results indicate that even if a control strategy destabilizes a fully synchronized state, its action on weakly synchronous dynamics may be quite opposite. This, perhaps, should not be surprising. The major idea behind desynchronizing algorithms like desynchronizing delayed feedback is that they are set up in such a way as to make the synchronous state unstable. However, neural synchrony in the human brain is far from being stable especially in pathological conditions; rather it varies in time and has a moderate average strength. Thus what defines synchrony strength in the system is not only the stability/instability properties of the synchronized state, but also the mechanisms pushing the system back to a synchronous state. The desynchronizing strategies such as delayed feedback are mostly concerned with destabilizing the synchronous state and do not address these other issues.

A number of neurological and psychiatric conditions have been associated with elevated levels of synchrony of neural oscillations [40,41]. Desynchronizing deep brain stimulation may have therapeutic potential for treatment of any conditions where excessive synchrony leads to pathological symptoms. However, as the current study suggests, to identify a viable desynchronization algorithm, one needs to test it in models with reasonably accurate reproduction of clinically relevant features of synchronized oscillatory activity.

**Figure Captions**

**Figure 1. The schematics of the model network with examples of synaptic connections between neurons.** Bars represent inhibitory synapses and arrows represent excitatory synapses. Delayed feedback stimulation $I_{stim}$ is modeled as an applied current to specified STN neurons.

**Figure 2. Stimulation setup for STN neurons.** STN LFP is first computed from synaptic currents and then band-pass filtered using a damped harmonic oscillator. Differential delay signal is then constructed from the filtered signal, bounded by nonlinear transformation, amplified and injected into the same neurons.

**Figure 3. Parameter plane with the number of principal components in the network without stimulation.** Dashed contours represent parameter values for which the model network synchronization dynamics is close to the experimental dynamics as analyzed in [26] for the weight parameter $w_1 = 0.3$. Simulations with feedback stimulation were performed for the parameter values inside the dotted rectangle. The filling of the circles specifies the number of principle components; red being synchronized dynamics and black being incoherent dynamics.

**Figure 4. Delayed feedback stimulation effect in the model network.** A) Desynchronization of strongly synchronous STN neurons with delayed feedback stimulation. B) Feedback stimulation leads to increased synchrony in the model network in the physiological intermittent regime. Half of model STN neurons are affected by stimulation. Electrodes are placed near the $5^{th}$ and $7^{th}$ STN neurons in the array and their membrane potentials (in mV) are shown in the top and bottom time traces in A) and B). Middle boxes contain voltage for the $5^{th}$ (red curve) and $7^{th}$ (blue curve) STN neurons together filtered to the beta-band. Stimulation is switched on at 2000 ms (indicated by vertical dashed line). Parameters are: $w_1 = 0.3$, $w_2 = 0$, $\tau_s = 50$ ms, $\tau = 50$ ms, $C_1 = 0.04$, $I_{app} = 5$, A) $g_{syn} = 1.3$, $K_f = 45$. B) $g_{syn} = 0.9$, $K_f = 35$. The mean period of the model network without stimulation is $T \approx 100$ ms.

**Figure 5. Change in the number of PCA components in the network with different feedback stimulation set-ups.** Positive change indicates dynamics less synchronous than pre-stimulation dynamics, negative change indicates more coherent dynamics. Two different spatial stimulation set-ups and two different weights are presented. A, B) 50% or 70% of STN neurons are directly affected by stimulation current, correspondingly. Electrodes are placed near the 5$^{th}$ and 7$^{th}$ STN neurons in the array. C, D) 50% or 70% of STN neurons are directly affected by stimulation current, correspondingly. Electrodes are placed near the 5$^{th}$, 6$^{th}$ and 7$^{th}$ STN neurons. Weight parameters are $w_1 = 0.3$, A, C) $w_2 = 0$; B, D) $w_2 = 0.1$. $I_{app} = 5$ in all simulations.

**Figure 6. Change in the number of PCA components in a network with feedback stimulation.** The spatial set-up of stimulation is different from that in Figure 5. A, B) 30% or 50% of STN neurons are directly affected by stimulation. Electrode is placed near the 5$^{th}$ STN neuron. C, D) 40% or 60% of STN neurons are directly affected by stimulation current, correspondingly. Electrodes are placed near the 5$^{th}$ and 6$^{th}$ STN neurons. E, F) 70% or 90% of STN neurons are directly affected by stimulation current, correspondingly. Electrodes are placed near the 5$^{th}$, 7$^{th}$ and 9$^{th}$ STN neurons. Weight parameters are $w_1 = 0.3$, A, C, E) $w_2 = 0$; B, D, F) $w_2 = 0.1$. $I_{app} = 5$ in all simulations.

**Figure 7. Maximum improvement in the number of principal components during stimulation.** The spatial electrode setups are the same as in Figure 5. While the desynchronizing action (filled circles) is consistent for the lower right corner (strongly correlated dynamics), it is very rare outside of that corner, for moderately synchronous (and more realistic) dynamics. Filled circles indicate desynchronizing action of stimulation of various efficiency (indicated by different colors). Empty circles indicate no desynchronization. Note that unlike Figure 5, here we consider the maximum improvement, so that it cannot be negative (it is always zero for zero stimulation strength). Dashed contours represent parameter values for which the model network synchronization dynamics is close to the experimental dynamics as analyzed in [26] for the weight parameter $w_1 = 0.3$. A, B) 50% or 70% of STN neurons are directly affected by the stimulation current. Electrodes are placed near the 5$^{th}$ and 7$^{th}$ STN neurons in the array. C, D) 50% or 70% of STN neurons are directly affected by stimulation current. Electrodes are placed

near the $5^{th}$, $6^{th}$ and $7^{th}$ STN neurons. Weight parameters are $w_1 = 0.3$, A, C) $w_2 = 0$; B, D) $w_2 = 0.1$.

**Figure 8. Maximum improvement in the number of principal components during stimulation.** The spatial electrode setups are the same as in Figure 6. These results are overall similar to those in Figure 7. Dashed contours represent parameter values for which the model network synchronization dynamics is close to the experimental dynamics as analyzed in [26] for the weight parameter $w_1 = 0.3$. A, B) 30% or 50% of STN neurons are directly affected by the stimulation current, correspondingly. Electrode is placed near the $5^{th}$ STN neuron in the array. C, D) 40% or 60% of STN neurons are directly affected by the stimulation current. Electrodes are placed near the $5^{th}$ and $6^{th}$ STN neurons. E, F) 70% or 90% of STN neurons are directly affected by the stimulation current. Electrodes are placed near the $5^{th}$, $7^{th}$ and $9^{th}$ STN neurons. Weight parameters are $w_1 = 0.3$, A, C, E) $w_2 = 0$; B, D, F) $w_2 = 0.1$.

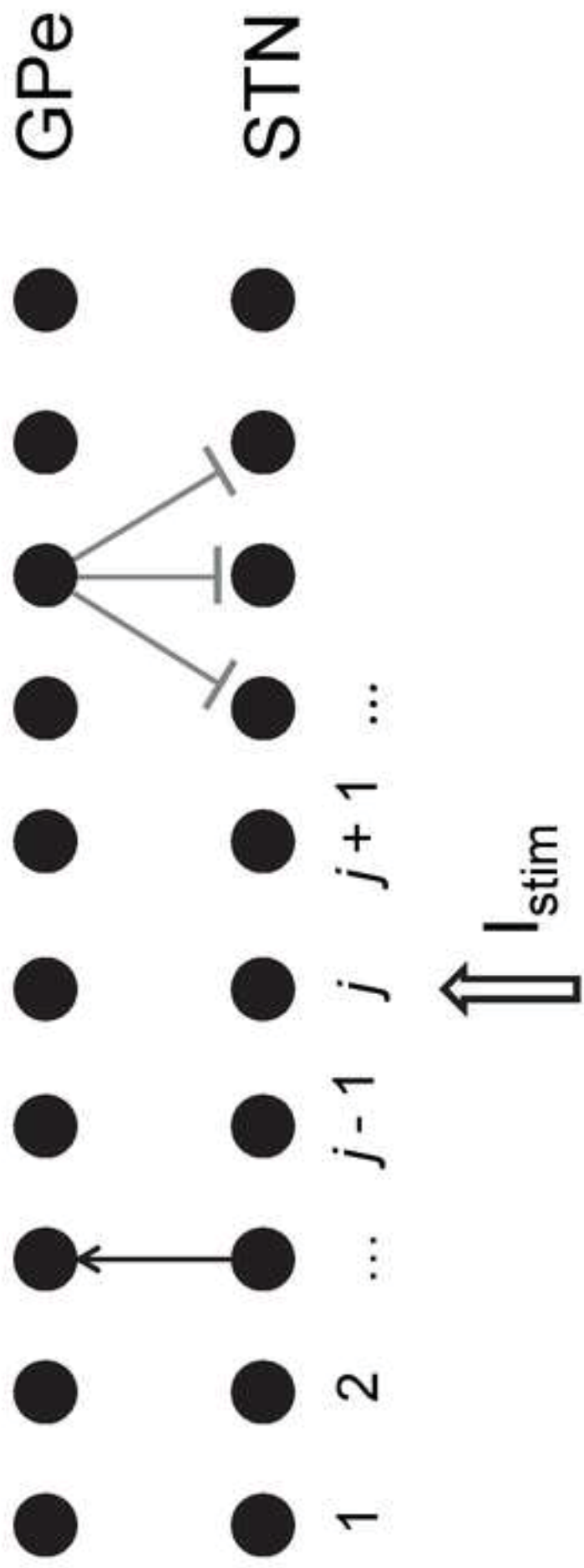

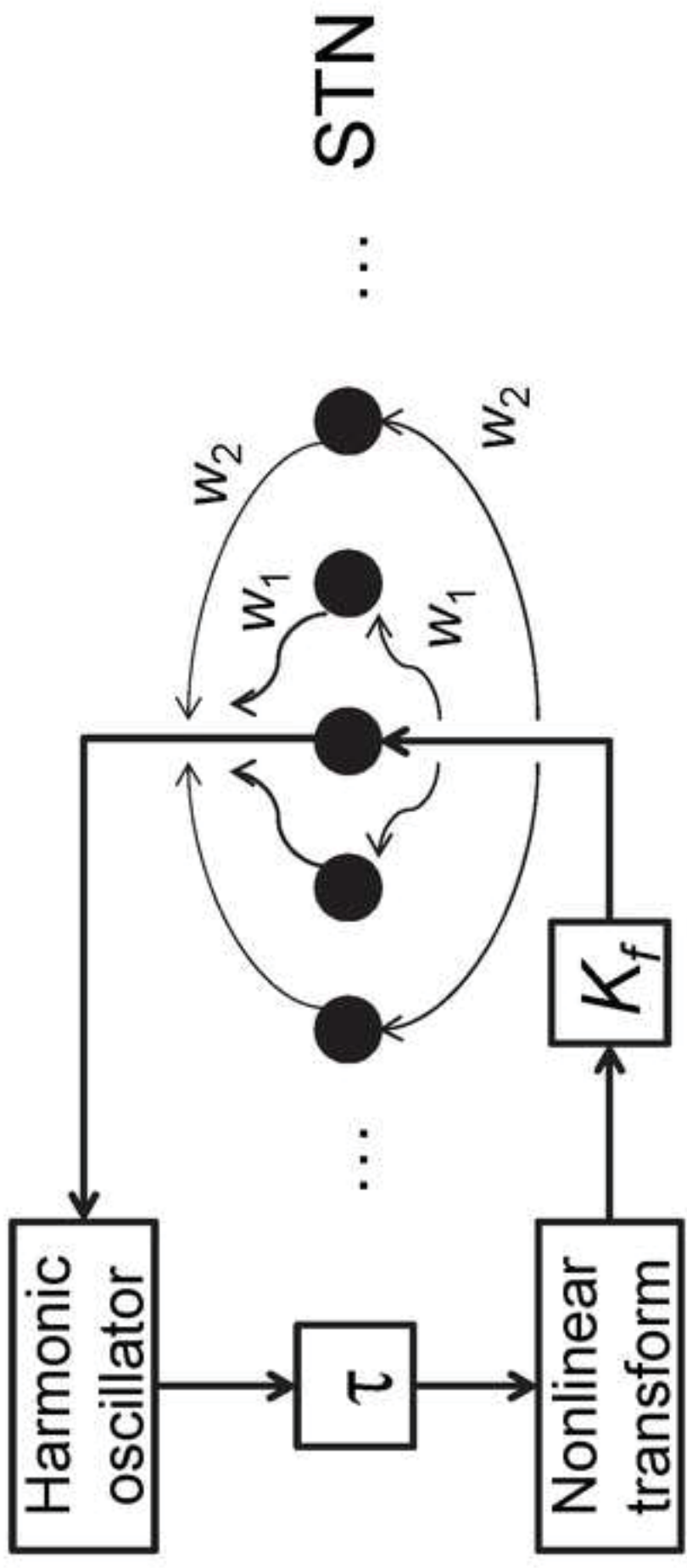

Figure2
Click here to download high resolution image

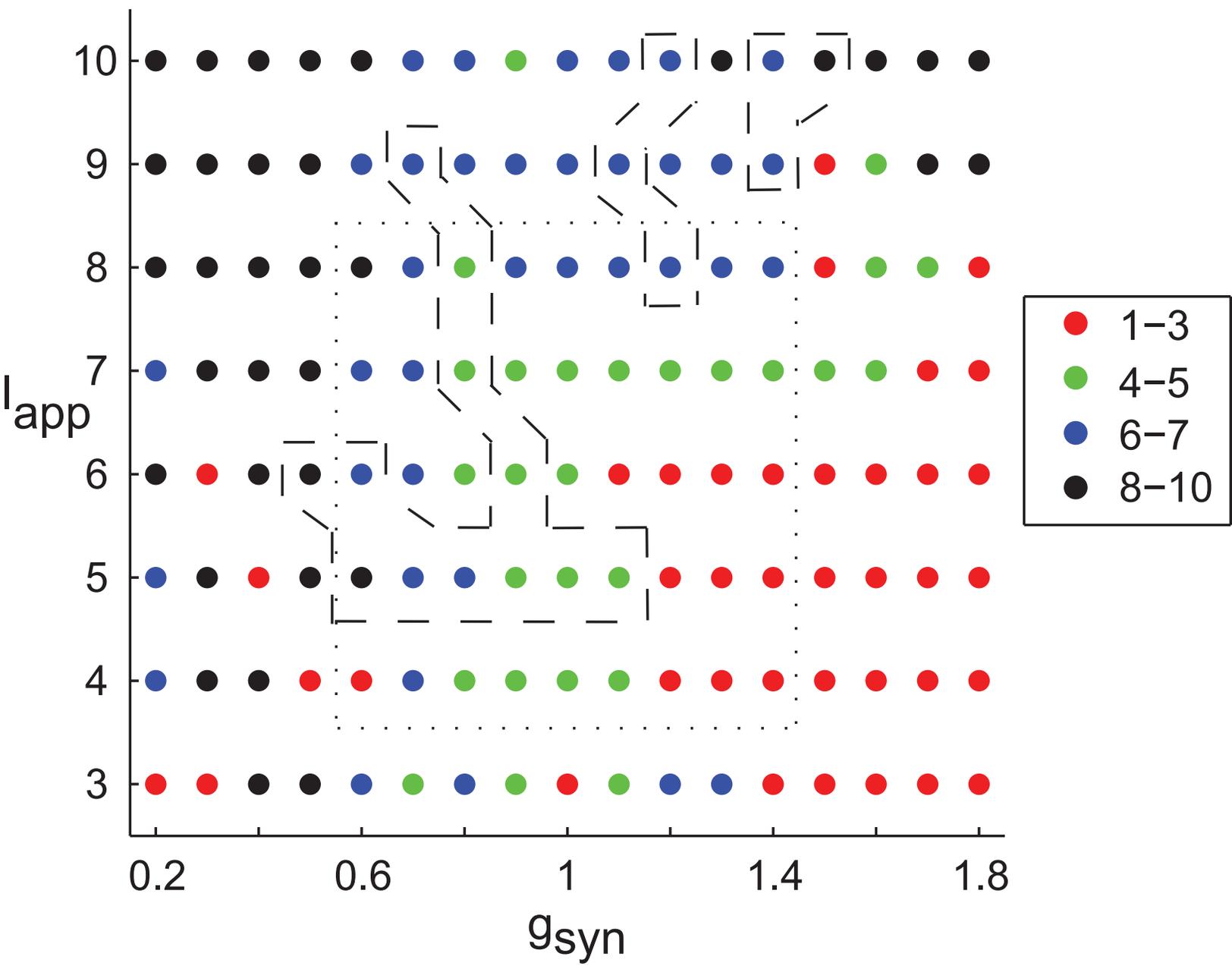

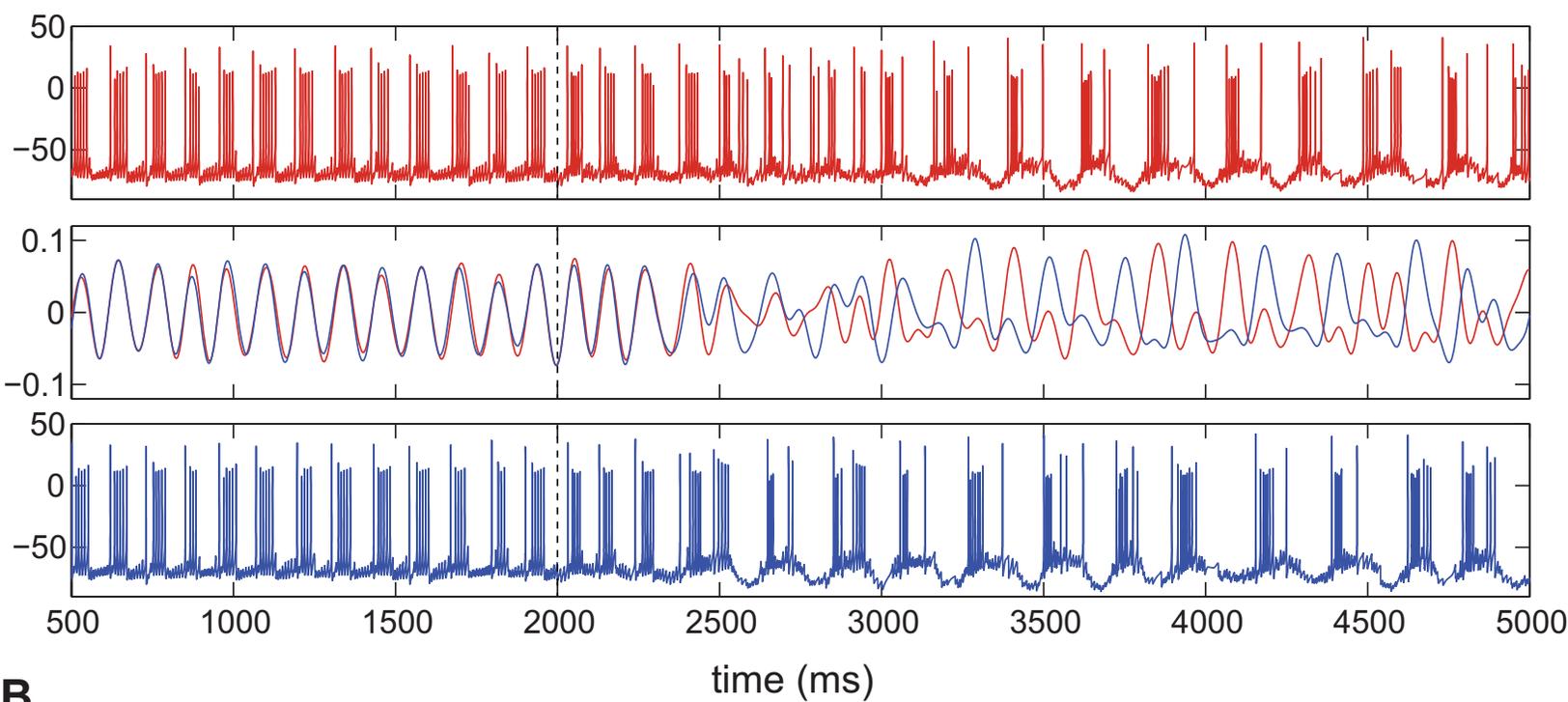

**A**

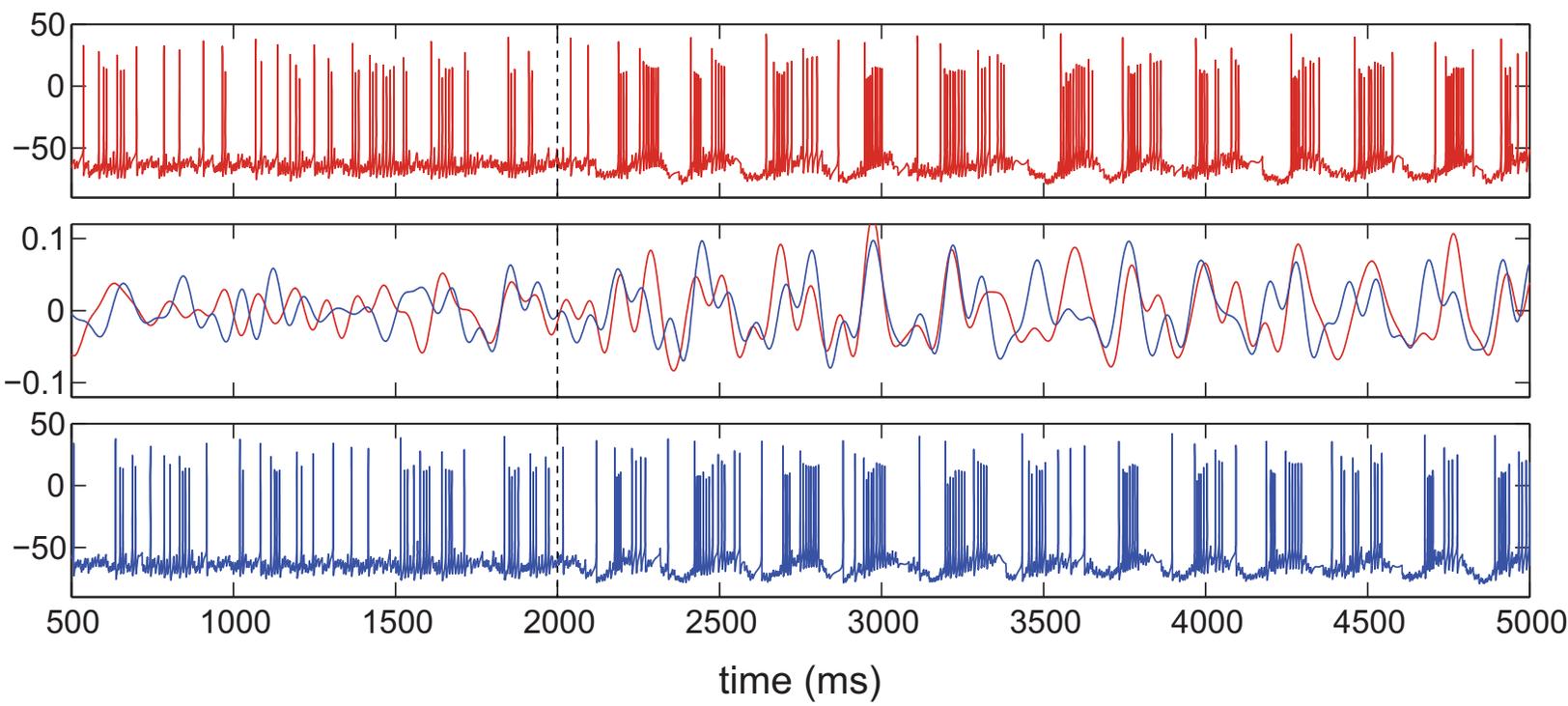

**B**

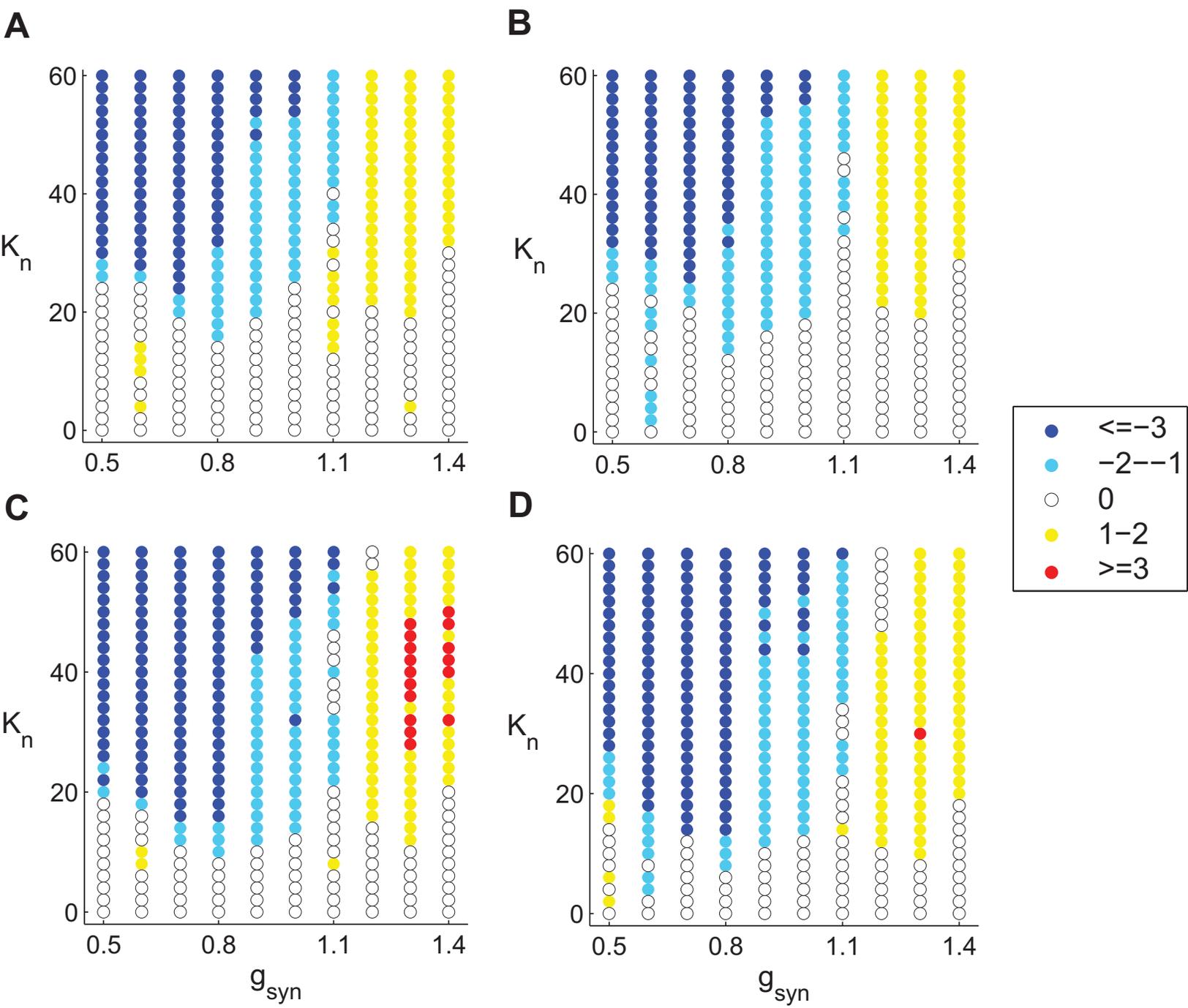

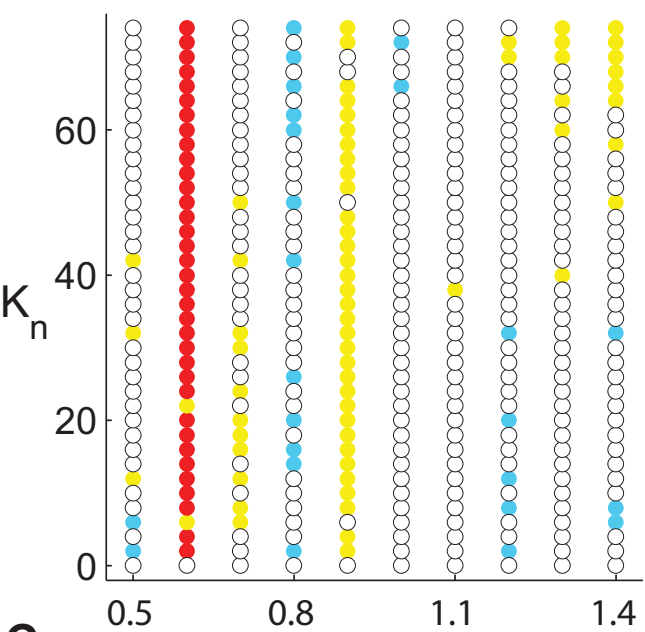
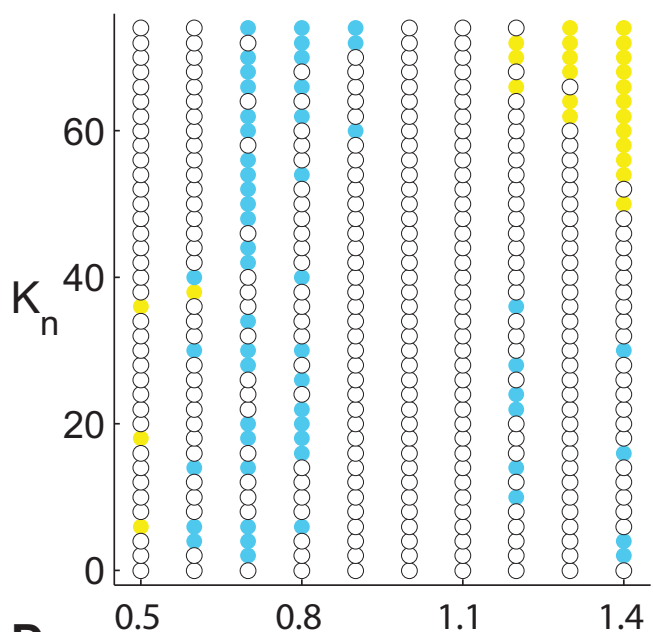
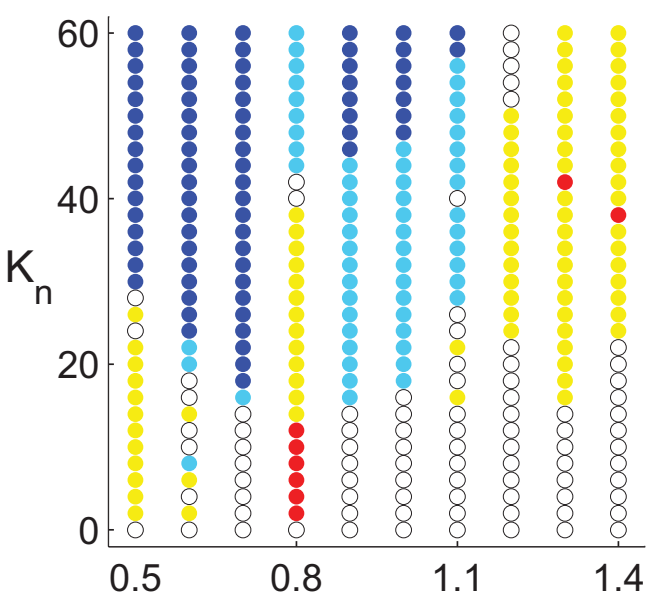
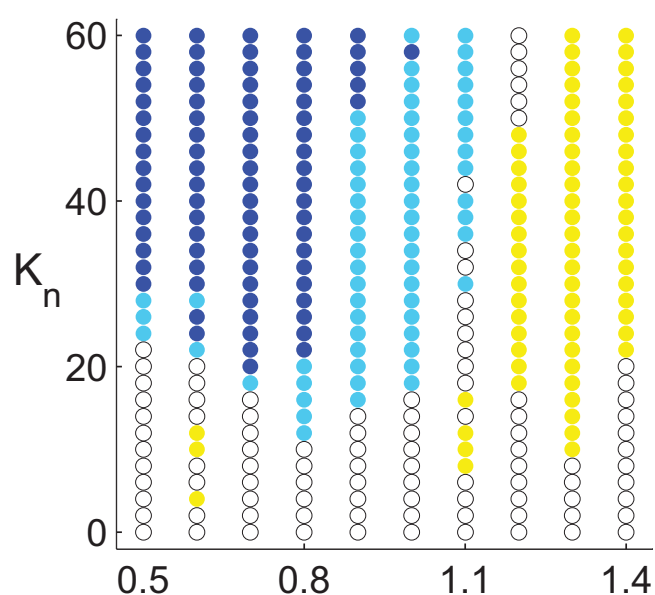
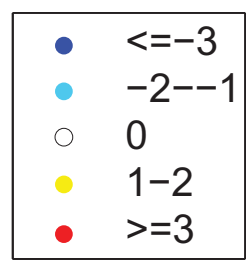
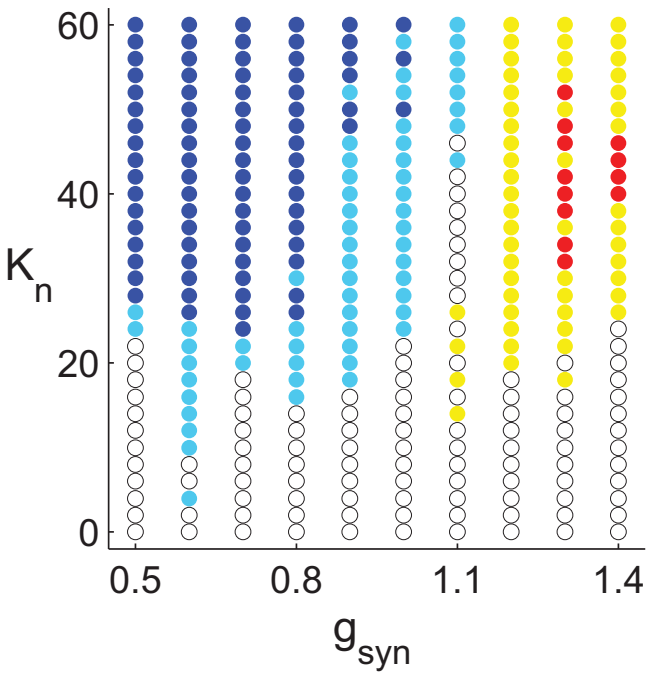
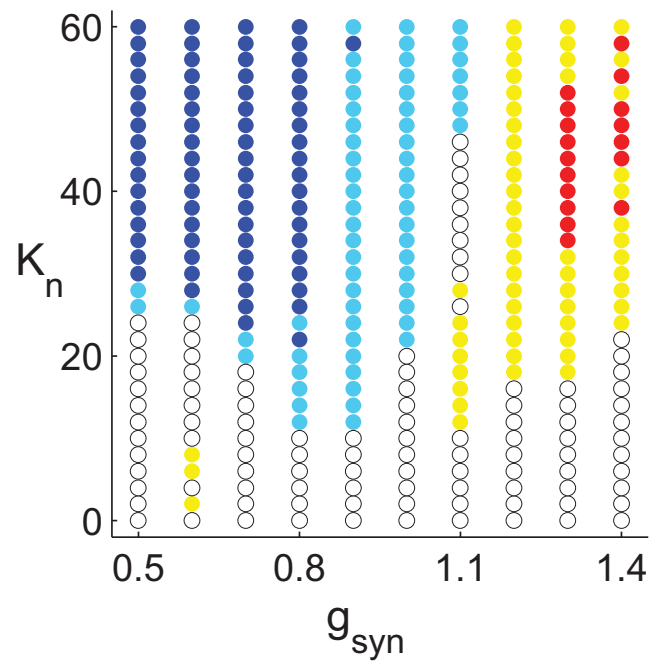

Figure 7
Click here to download Figure: Fig7.eps

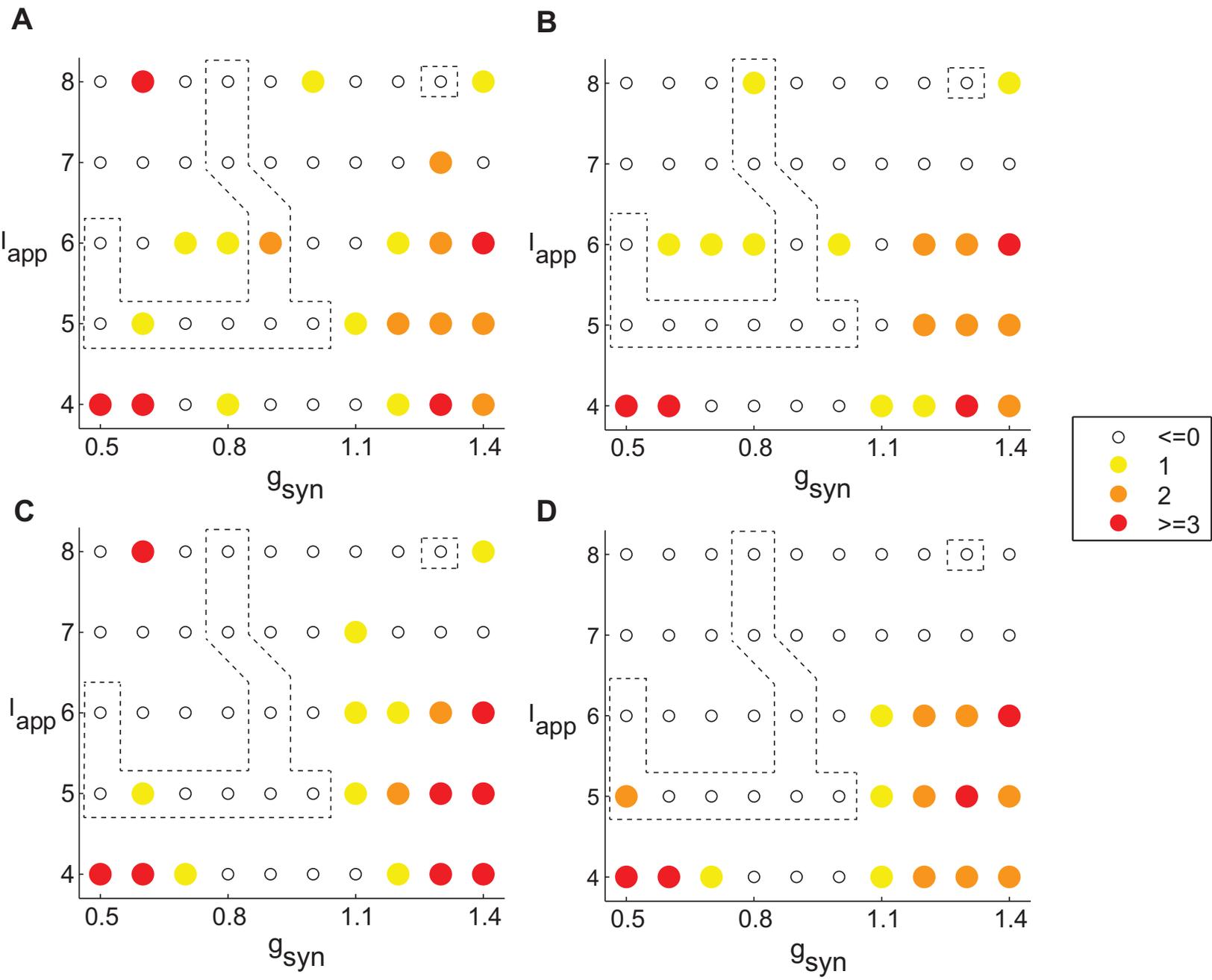

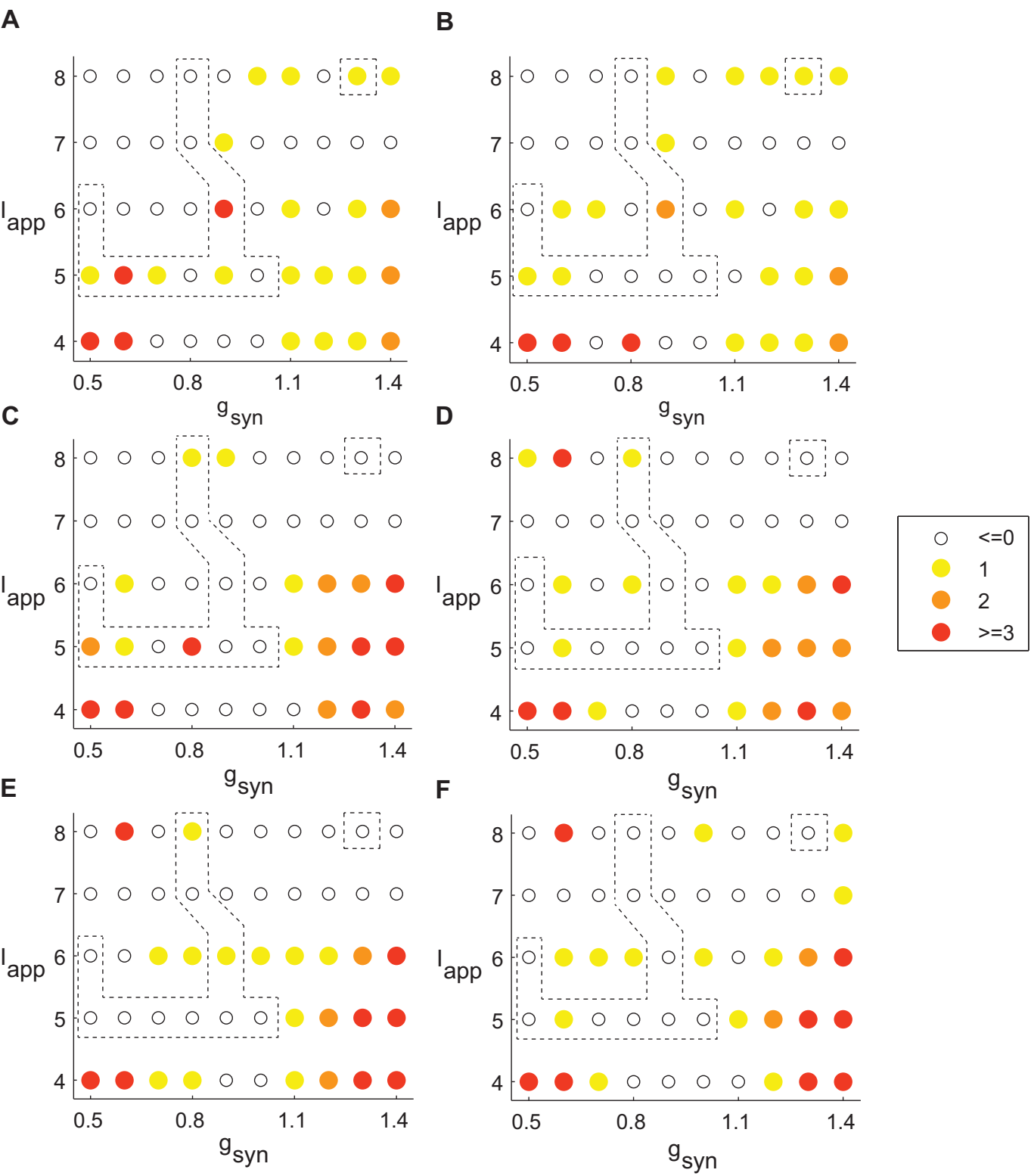